\documentclass[mathleft]{an}
\usepackage{graphicx}
\usepackage{times}
\begin{document}

% The following seven commands are intended for editorial usage and should be ignored by
% the author(s).
\Pagespan{789}{}% Document's page range.
% If second parameter is left empty, the last page is computed automatically.
\Yearpublication{2006}%
\Yearsubmission{2005}%
\Month{11}%
\Volume{999}%
\Issue{88}%
% \DOI{This.is/not.aDOI}%

\title{Ultraluminous X-ray pulsar: accreting magnetar?}

\author{H. Tong\thanks{Corresponding author:
  \email{tonghao@xao.ac.cn}\newline}
%Example
%for footnote, note the usage of the \texttt{fnmsep}
%command as separator between institute number and footnote mark}
}
\titlerunning{ULX pulsar as accreting magnetar}
\authorrunning{Tong}
\institute{Xinjiang Astronomical Observatory, Chinese Academy of Sciences, Urumqi 830011, China}

\received{30 May 2005}
\accepted{11 Nov 2005}
\publonline{later}


\abstract{Magnetars are a special kind of neutron stars. There may also be accreting magnetars. From the studies of isolated magnetars, 
it is known that a neutron star with a strong dipole field only is not a magnetar. Super-slow X-ray pulsars may just be 
accreting high magnetic field neutron stars. The ultra-luminous X-ray pulsar NuSTAR J095551+6940.8 may be an accreting magnetar. 
It may be an accreting low magnetic field magnetar with multipole field of $10^{14} \,\rm G$ and dipole field of $10^{12} \,\rm G$. 
This point of view is consistent with the study of isolated magnetars.  
An ultra-luminous X-ray pulsar phase in the binary evolution may result in massive millisecond pulsars.}

\maketitle

\section{Introduction}

Pulsars are rotating neutron stars. They were first discovered in 1967 (Hewish et al. 1986). 
With the advance of X-ray astronomy, accreting neutron stars were discovered in the 1970s (Giacconi et al. 1971).
They manifest themselves as accreting X-ray pulsars. In 1982, millisecond pulsars were discovered (Backer et al. 1982). 
They are commonly assumed to be recycled neutron stars via low mass X-ray binaries (Alpar et al. 1982). 
In this way, normal pulsars, accreting neutron stars, and millisecond pulsars are linked together. 
In the 1990s, the idea of strongly magnetised neutron stars (i.e. magnetar) were theoretically proposed 
and observationally discovered (Duncan \& Thompson 1992; Kouveliotou et al. 1998). 
Since magnetars are just a special kind of isolated neutron stars, it is natural to ask 
``where are accreting magnetars?'' (Woods \& Thompson 2006). 

According to the general results of accretion power, the accreted matter onto magnetars 
would release their gravitational energy mainly in X-rays. The typical photon energy would be in the range of 
$1\,\rm keV$ to $50\,\rm MeV$ (Frank et al. 2002). Therefore, accreting magnetars are expected to be some kind 
of X-ray pulsars. Currently, four kind of X-ray pulsars are known. 
\begin{enumerate}
\item Rotation-power X-ray pulsars. The X-ray emissions of the Crab pulsar and the Vela pulsar etc are assumed to be originated 
from their rotational energy (Becker 2009). 

\item Accretion powered X-ray pulsars. They are accreting neutron stars in X-ray binaries (Bhattacharya \& van den Heuvel 1991).

\item Magnetars. Observationally, anomalous X-ray pulsars and soft gamma-ray repeaters are thought to be magnetar candidates
(Olausen \& Kaspi 2014; Mereghetti et al. 2015; Tong \& Xu 2014). The persistent and burst energy of anomalous X-ray pulsars
and soft gamma-ray repeaters may originate from the neutron star's magnetic energy. 

\item If the above three kinds of energy reservoir are all out of reach, then the central neutron star can still emit X-rays due to its relic thermal energy (Page et al. 2006).  The thermal X-ray emission may also present in other types of X-ray pulsars, e.g. rotation-powered X-ray pulsar. X-ray dim isolated neutron stars (Turolla 2009) and central compact objects (Gotthelf et al. 2013) have very nice thermal X-ray spectra. 
 
\end{enumerate}
Then both theory and observations should tell the differences between accreting magnetars and the four kinds of X-ray pulsars known at present. 

For accreting compact objects, including accreting magnetars, the maximum luminosity for steady spherical accretion is (i.e. Eddington luminosity, Frank et al. 2002):
\begin{equation}
L_{\rm Edd} = 1.3\times 10^{38} M_{1} \,\rm erg \,s^{-1},
\end{equation}
where $M_{1}$ is the mass of the central object in units of solar masses. For a neutron star with whose radius is about 
$10^6 \,\rm cm$, its effective temperature is about $2\times 10^7 \,\rm K$ when radiating at the Eddington luminosity. 
If the central neutron star is radiating at hundreds or thousands of times super-Eddington luminosity, then its radiation may 
mainly in the hard X-ray range. This is why the giant flares of magnetars and ultra-luminous X-ray pulsar are mainly observed 
using hard X-ray telescopes (Mereghetti 2008 and references therein; Bachetti et al. 2014). For the ultra-luminous X-ray pulsar, 
it can also be observed in the soft X-ray range. However, its timing properties can only be studied using the {\it NuSTAR} hard X-ray 
telescope (Bachetti et al. 2014). 

Some basic knowledge of isolated magnetars is provided in Section 2.  
The case of super-slow X-ray pulsars is discussed in Section 3. 
The discovery of ultraluminous X-ray pulsar and its explanations in the accreting magnetar model are presented 
in Section 4. Discussion on current observations and models is given in Section 5. 
Conclusion of this paper is provided in Section 6. 

\section{Basic physics of isolated magnetars}

\begin{figure}[!htbp]
\centering
 \includegraphics[width=0.45\textwidth]{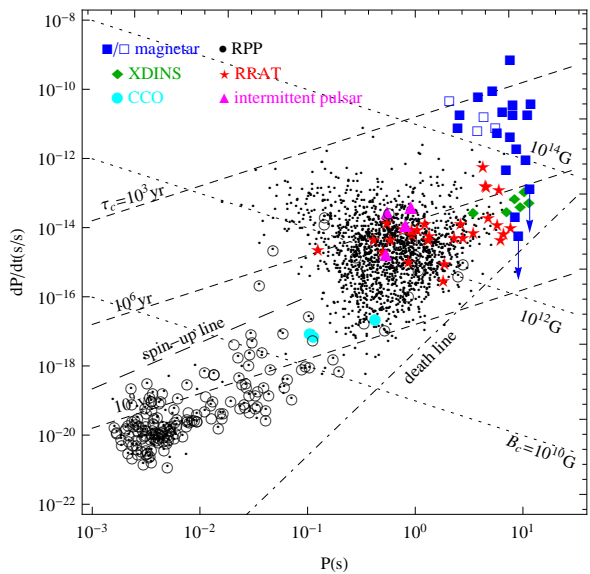}
\caption{Period and period-derivative diagram of various pulsar-like objects.
Black dots are normal rotation-powered pulsars, blue squares are magnetars (empty squares are radio-loud magnetars). 
Dot with a circle means the pulsar is in a binary system. 
Updated from figure 1 in Tong \& Wang (2014).}
\label{fig_PPdot}
\end{figure}

The period and period-derivative diagram of various pulsar-like objects is shown in figure \ref{fig_PPdot}. 
From the measured period and period derivative, the characteristic magnetic field and characteristic age of 
pulsars and magnetars can be obtained by assuming magnetic dipole braking in vacuum (Tong 2015a). 
For magnetars with typical period of 10 seconds, and period derivative
of $10^{-11}$, their characteristic magnetic field is 
\begin{equation}
B_{\rm c} =3.2 \times 10^{19} \sqrt{P\,\dot{P}} \sim 3\times 10^{14} \,\rm G, 
\end{equation}
and characteristic age
\begin{equation}
\tau_{\rm c} =\frac{P}{2\dot{P}} \sim 2\times 10^{4} \,\rm yr. 
\end{equation}
This forms the basic belief that magnetars are young neutron stars with strong magnetic field (Mereghetti 2008). 
However, this basic picture is too simple to be true. Some pulsars can have magnetic field as high as $10^{14} \,\rm G$
while not showing significant magnetar activities (Ng \& Kaspi 2011). The discovery of low magnetic field magnetars 
(with characteristic magnetic field less than $7.5\times 10^{12} \,\rm G$, Rea et al. 2010) further shows that 
the dipole magnetic field is not the key ingredient of magnetars. Magnetars may be neutron stars with strong multipole field only
(Tong et al. 2013 and references therein). The strong multipole field is responsible for the bursts (including giant flares),
super-Eddington luminosity during bursts, persistent emissions, and spin down of magnetars etc. 

\begin{itemize}
\item \textbf{Giant flares} Three giant flares are observed from three different magnetars (Mereghetti 2008 and references therein).
The giant flare is made up of a spike and a pulsating tail (lasts for hundreds of seconds). During the pulsating tail, the X-ray luminosity 
can be as high as $10^{42} \,\rm erg \, s^{-1}$ which is about $10^4$ times the Eddington luminosity. In the magnetar model, the strong multipole field as high as $10^{15} \,\rm G$ provides the energy budget of giant flare (Thompson \& Duncan 1995).
For a $B \sim 10^{15} \,\rm G$ multipole field in the vicinity of the neutron star surface, its typical spatial scale may be as that of the neutron star radius about $R \sim 10^6 \,\rm cm$. Then the total magnetic energy will be 
\begin{equation}
E_{\rm B} \sim \frac{B^2}{8\pi} \,R^3 \sim 4\times 10^{46} \,\rm erg. 
\end{equation}
If a significant amount of this energy can be release is a catastrophic way, it may corresponds to the giant flare of magnetars (Yu \& Huang 2013). At the same time, if the magnetic energy can be release gradually this may correspond to its persistent state X-ray luminosity (Thompson \& Duncan 1996). For a typical time scale of $\tau \sim 10^{5} \,\rm yr$ which is the typical age of magnetars, the X-ray luminosity is 
\begin{equation}
L_{\rm x} \sim \frac{E_{\rm B}}{\tau} \sim 10^{34} \,\rm erg \,s^{-1}. 
\end{equation}
Furthermore, the magnetic field will also suppress the scattering cross section between electrons and photons (Paczynski 1992). The traditional Eddington luminosity uses the Thomson cross section and does not consider the effect of strong magnetic field (Frank et al. 2002). In the presence of strong magnetic field, the corresponding critical luminosity is (Paczynski 1992):
\begin{equation}\label{LEddingtonB}
\frac{L_{\rm cr}}{L_{\rm Edd}} \approx 2\times \left( \frac{B}{10^{12 \,\rm G}} \right)^{4/3}.
\end{equation} 
For a magnetic field strength of $10^{15} \,\rm G$, the corresponding critical luminosity can be $10^{4}$ times the traditional Eddington luminosity. 
This may explain the super-Eddington pulsating tail in magnetar giant flares 
(Paczynski 1992; Mereghetti 2008). 
If a magnetar is accreting from a binary companion, its luminosity can also be hundreds or thousands times the traditional Eddington
luminosity. This inspired the accreting magnetar model for the ultra-luminous X-ray pulsar (Section 4 below). 

\item \textbf {Timing behaviors} 
Compared with normal rotation-powered pulsars, magnetars are always variable (Rea \& Esposito 2011). They can have burst, and outbursts. During these active episode, the magnetar's timing behavior also changes (e.g. Archibald et al. 2015 and references therein; Tong \& Xu 2014). 
The characteristic magnetic field of the low magnetic field magnetar SGR 0418$+$5729 may be related to its small magnetic inclination angle (Tong \& Xu 2012). The decreasing period derivative of the second low magnetic field magnetar Swift J1822.3$-$1606 may be due to its decreasing particle wind (Tong \& Xu 2013). The anti-correlation between timing and X-ray luminosity of the galactic centre magnetar SGR J1745$-$2900 may due to its changing geometry (Tong 2015b). 
The so called anti-glitch
of magnetar 1E 2259$+$586 may simply be a period of enhanced spin down rate (Tong 2014). All these timing behaviors of magnetars is understandable in the wind braking model of pulsars (Tong et al. 2013). 

\item \textbf {Spectra properties} Magnetars can be detected in the hard X-ray range (Gotz et al. 2006). There may be a cutoff around $1\,\rm MeV$ in the high energy spectrum of magnetars (Abdo et al. 2010). 
A possible cutoff at about $130\,\rm keV$ is found in magnetar 4U 0142$+$61 using nine years of {\it INTEGRAL} data (Wang et al. 2014). A cutoff at $130\,\rm keV$ will rule out ultra-relativistic emission models for the hard X-rays. The putative electrons should be at most mildly relativistic. Compared with the hard X-ray emission of rotation-powered pulsars and accretion-powered pulsars (Mereghetti 2015, figure 3 there; Wang et al. 2014), magnetar's hard X-ray emission will dominate above $100\,\rm keV$. Therefore, for an accreting magnetar, a hard X-ray tail above $100\,\rm keV$ may present. This may be another signature of accreting magnetars. 
\end{itemize}

In summary, the study of isolated magnetars tells us that the difference between magnetars and normal rotation-powered pulsars is their strong multipole field. It is not their positions on the period and period diagram which is determined mainly the magnetospheric torque. 
Therefore, if people want to find accreting magnetars, the signature of strong multipole field in accreting systems are expected (Tong \& Wang 2014):  (1) magnetar-like bursts. The magnetar burst is different from type I or type II burst in accreting systems (Gavriil et al. 2002). (2) a hard X-ray tail above $100\,\rm keV$. China's future hard X-ray modulation telescope (HXMT) may contribute on these aspects (Guo et al. 2015a). 

\section{Super-slow X-ray pulsars are not accreting magnetars}

Super-slow X-ray pulsars are a special kind of accreting neutron stars (Tong \& Wang 2014): (1) They have period longer than 
$10^{3} \,\rm s$; (2) large period derivatives $10^{-7}-10^{-6}$ which may be either positive or negative; (3) their X-ray luminosity ($1-100\,\rm keV$) is about $10^{34}-10^{37} \,\rm erg \, s^{-1}$ which may be accretion of stellar wind from the neutron star's high mass companion. Theoretically, if their long pulsation period is equal to the accretion equilibrium period, the corresponding magnetic field required will be about $10^{15} \,\rm G$. Their large period derivative also require a very high magnetic field. 
Therefore, super-slow X-ray pulsars are often assumed to be accreting magnetars. However, the magnetic field determined from the timing behavior is the large scale dipole field and it also depends on the accretion torque\footnote{Using a different torque, a strong dipole field may not be required (Shakura et al. 2012).}. More importantly, even if the central neutron star has a very high dipole field, it is at most an accreting high magnetic field neutron star. In order to say that an accreting magnetar is seen evidences of strong multipole field are required. Considering that super-slow X-ray pulsars may have very high dipole field, it is also possible that some of them may also have very strong multipole field. Therefore, they are candidates for searching of magnetic activities in future observations. A comparison between isolated magnetars and super-slow X-ray pulsars are summarised in Table 1 in Tong \& Wang (2014) . 

\section{Ultraluminous X-ray pulsar and the accreting magnetar model}

Bachetti et al. (2014) discovered that NuSTAR J095551+6940.8 is an ultraluminous X-ray pulsar with isotropic X-ray luminosity about 
$10^{40} \,\rm erg \,s^{-1}$. One previously identified ulltraluminous X-ray source M82 X-2 may be the soft X-ray counterpart of this ultra luminous X-ray pulsar (Feng \& Kaaret 2007; Kong et al. 2007; Bachetti et al. 2014).  The pulsation period of the pulsar is $1.37\,\rm s$. At the same time, the pulsar is spinning up at an rate about $\dot{P} \approx -2\times 10^{-10}$. The super-Eddington luminosity and the spin-up trend pose challenges to the accretion theory of neutron stars. 

From previous study of isolated magnetars, an aged magnetar is more like to be a low magnetic field magnetar (Rea et al. 2010; Turolla et al. 2011).
For an accreting magnetar, the neutron star may be relatively old with age about $10^6 \,\rm yr$. 
Then it is an accreting low magnetic field magnetar.  
Tong (2015c) pointed that NuSTAR J095551+6940.8 may be an accreting low magnetic field magnetar. The magnetar strength multipole field (about $10^{14} \,\rm G$) is responsible for the super-Eddington X-ray luminosity (cf. equation(\ref{LEddingtonB})). The much lower large scale dipole field
(e.g. about $10^{12} \,\rm G$) is responsible for the interaction between the accreting flow and the central neutron star. In this case, the spin-up rate is understandable even in the traditional form of accretion torque. 

Far away from the neutron star, the accretion flow may of a disk form if the accreted material posses enough angular momentum. Closer to the neutron star the effect of the central star magnetic field is stronger. The Alfv$\rm\acute{e}$n radius is defined as where ram pressure of the accretion flow equals the local magnetic energy density (Shapiro \& Teukolsky 1983). Inside the Alfv$\rm\acute{e}$n radius, the accretion flow is controlled by the magnetic field. 
The Alfv$\rm\acute{e}$n radius depends on the geometry of the accretion flow. 
It is approximately (Shapior \& Teukolsky 1983; Lai 2014)
\begin{equation}
 R_{\rm A} = 3.2\times 10^8 \mu_{30}^{4/7} M_{1}^{-1/7} \dot{M}_{\rm 17}^{-2/17} \,\rm cm,
\end{equation}
where $\mu_{30}$ is dipole magnetic moment of the central neutron star, $\dot{M}_{17}$ is the mass accretion rate in units of $10^{17} \,\rm g \,s^{-1}$.The light cylinder radius is defined as the radius where the corotational velocity 
is equal to the speed of light
\begin{equation}
 R_{\rm lc} = \frac{P c}{2\pi} = 6.5\times 10^9 \,\rm cm,
\end{equation}
where in the last step the period of the ultra-luminous X-ray pulsar is substituted. 
When the Alfv$\rm\acute{e}$n radius is less than the light cylinder radius, the accretion flow may interact with the central neutron star. 
The rcorotation radius is defined as the radius where the corotational velocity equals the local Keplerian velocity
\begin{equation}
 R_{\rm co} = \left( \frac{G M P^2}{4\pi^2} \right)^{1/3} = 1.8\times 10^8 M_{1}^{1/3} \,\rm cm,
\end{equation}
the period of the ultra-luminous X-ray pulsar is substituted. 
When the Alfv$\rm\acute{e}$n radius is less than the corotation radius, large scale accretion is possible. Otherwise, the accretion flow is stopped 
by the centrifugal barrier (i.e. propeller, Illarionov \& Sunyaev 1975). Accretion equilibrium corresponds to when the corotational radius is equal 
to the Alfv$\rm\acute{e}$n radius: $R_{\rm co} = R_{\rm A}$ (Lai 2014). 
The dipole magnetic field can be inferred from
the spin-up trend of the ultra-luminous X-ray pulsar (Tong 2015c). The angular momentum carried by the accreted matter onto the neutron star is 
$\dot{M}_{\rm acc}\sqrt{G M R_{\rm A}}$. The spin-up rate of the central neutron star is determined by
\begin{equation}
 -2\pi I \frac{\dot{P}}{P^2} = \dot{M}_{\rm acc} \sqrt{G M R_{\rm A}},
\end{equation}
where $I$ is the star's moment of inertia. The mass accretion rate can be inferred from the X-ray luminosity considering possible beaming effect.
For a two solar mass neutron star, with a beaming factor of 0.2, the dipole magnetic field is about (Tong 2015c)
\begin{equation}
 B_{\rm p} = 1.6\times 10^{12} R_6 L_{\rm iso,40}^{-3} \,\rm G, 
\end{equation}
where $R_6$ is the neutron star radius in units of $10^6 \,\rm cm$, $L_{\rm iso,40}$ is the isotropic luminosity in units of $10^{40} \,\rm erg \,s^{-1}$. 

Several other interesting aspects can also be obtained (Tong 2015c): (1) M82 X-2 is a transient because it may switch between accretion phase and propeller phase. (2) The soft X-ray excess in M82 X-2 may due to emission from the accretion disk. Future more accurate determination of the inner disk radius will constrain different models. (3) The theoretical pulsation period of accreting magnetars may be very wide which ranges from $0.1\,\rm s$ to $10^{3} \,\rm s$. (4) Up to now, three signatures of accreting magnetar are available: magnetar-like bursts, a hard X-ray tail, and an ultraluminous X-ray pulsar. 

\section{Discussion}

Both the radiation and rotation behaviors of NuSTAR J095551+6940.8 may due to effects of magnetic field. According to the current 
models, there are at least three ways to constrain the magnetic field strength\footnote{Measurement of cyclotron lines in the future will be very useful.}:
\begin{enumerate}
 \item Maximum luminosity in the presence of a strong magnetic field. The observed isotropic X-ray luminosity $L_{\rm x} \approx 10^{40} \,\rm erg \,s^{-1}$
 should be smaller than the maximum accretion luminosity onto magnetic neutron stars. The maximum luminosity depends on the magnetic field strength and has 
 been modeled by several authors (Paczynski 1992; Mushtukov et al. 2015). 
 
 \item Equilibrium state. The X-ray luminosity may switch from accretion phase to propeller phase (Tong 2015c; Dall'Osso et al. 2015; Tsygankov et al. 2015). 
 The pulsar may be in spin equilibrium (Eksi et al. 2015). Both of these two cases correspond to accretion equilibrium state which is characteristized 
 by equaling the corotation radius and Alfv$\acute{e}$n radius. 
 
 \item Modeling the rotational evolution. The spin-up trend of NuSTAR J095551+6940.8 depends on the accretion torque. By assuming some kind accretion torque, the magnetic field strength can be obtained (Bachetti et al. 2014; Eksi et al. 2015; Tong 2015c; Dall'Osso et al. 2015). 
 
\end{enumerate} 
During the calculations, there are various uncertainties: the exact value of the X-ray luminosity (it changes with time); the beaming factor;
the form of accretion torque; and various factors of 0.5 etc. Therefore, different authors may give different results of magnetic field strength. 
The exact value of magnetic field strength may range from $10^{12} \,\rm G$ to $10^{14} \,\rm G$. It seems that the behavior of NuSTAR J095551+6940.8 can be understood in a very wide parameter space at present. Future more observations of more sources will make clear the theoretical uncertainties.

The consequences of an ultra-luminous X-ray pulsar phase will be significant, especially for the star mass, magnetic field, and rotation period.
For an ultra-luminous X-ray pulsar phase with luminosity $L_{\rm x} \sim 10^{40} \,\rm erg\,s^{-1}$, the corresponding mass accretion rate 
will be $\dot{M} \sim 10^{20} \,\rm g \,s^{-1}$. For a neutron star with initial mass of $1.4\,\rm M_{\odot}$, it will grow to $2 \,\rm M_{\odot}$ in about
$4\times 10^5 \,\rm yr$. Its mass will be $2.5\,\rm M_{\odot}$ in about $7\times 10^5 \,\rm yr$. Therefore, the ultra-luminous X-ray pulsar phase is very efficient in producing massive neutron stars if this phase can lasts about $10^6\,\rm yr$ 
(Fragos et al. 2015; Wiktorowicz et al. 2015; Guo et al. 2015b)\footnote{The massive pulsar itself may account of the super-Eddington luminosity of the ultra-luminous X-ray pulsar (Guo et al. 2015b)}. 
In analogy with classical accreting neutron stars, the star's magnetic field may be reduced by the accretion matter 
(equation (20) in Zhang \& Kojima 2006 and references therein). For an initial magnetar strength dipole field of $10^{14} \,\rm G$, 
after accreting about $0.6\,\rm M_{\odot}$, the final magnetic field will be about $10^9\,\rm G$--$10^{11} \,\rm G$. Assuming spin equilibrium, 
the star's final rotation period will be about $0.3\,\rm ms$--$14\,\rm ms$ (not considering various instabilities which may limit the minimum spin period). Therefore, the ultra-luminous X-ray pulsar phase may be another 
channel to form millisecond pulsars\footnote{Kluzniak \& Lasota (2015) also proposed that ultra-luminous X-ray pulsars may evolve to millisecond pulsars. 
The conclusion there is from a very different point of view. } (Pan et al. in preparation). If sub-millisecond pulsars exist, they can be formed 
via the ultra-luminous X-ray pulsar phase. 

\section{Conclusion}

From the studies of isolated magnetars, it is known that the difference between normal pulsars and magnetars is their multipole field. 
A neutron star with a strong dipole field only is not a magnetar. Super-slow X-ray pulsars are probably not accreting magnetars. They may be 
accreting high magnetic field neutron stars. The ultra-luminous X-ray pulsar NuSTAR J095551+6940.8 may be an accreting magnetar. 
Its radiation and rotation behaviors can be understood in a wide parameter space. However, the exact value of its magnetic field strength 
is still uncertain. It may be an accreting low magnetic field magnetar with multipole field of $10^{14} \,\rm G$ and dipole field of $10^{12} \,\rm G$. 
This point of view is consistent with the study of isolated magnetars which tells that an aged magnetar is more likely to be a low magnetic field magnetar. An ultra-luminous X-ray pulsar phase in the binary evolution may result
in massive millisecond pulsars. Three signature of accreting magnetars are known at present: (1) magnetar-like bursts; (2) a hard X-ray tail; and (3) 
an ultra-luminous X-ray pulsar. 
Future hard X-ray telescopes (e.g. the HXMT telescope in China) may contribute to these aspects.

\acknowledgements
The author would like to thank W. Wang, Y. J. Guo, R. X. Xu, Y. Y. Pan, C. M. Zhang, C. H. Lee, and S. Popov for discussions which deepened my understanding of accreting neutron stars. 
H.Tong is supported by Xinjiang Bairen project, West Light Foundation of CAS (LHXZ201201), Qing Cu Hui of CAS, and 973 Program (2015CB857100).

%\newpage

\end{document}